\documentclass[12pt]{article}

\usepackage{sbc-template}
\usepackage{graphicx,url}
\usepackage[brazil]{babel}
\usepackage[utf8]{inputenc}

\usepackage{makecell}
\usepackage{arydshln}

\usepackage{amsmath}
\usepackage[dvipsnames]{xcolor}
\usepackage{amssymb}
\usepackage{soul}
\title{LLM-Based Intelligent Agents for Music Recommendation: A Comparison with Classical Content-Based Filtering}

\author{
Ronald Carvalho Boadana\inst{1}, Ademir Guimarães da Costa Junior\inst{1} \\
Ricardo Rios\inst{1}, Fábio Santos da Silva\inst{1}
}

\address{Escola Superior de Tecnologia - Universidade do Estado do Amazonas (UEA) \\ Manaus - AM - Brasil
    \email{\{rcb.eng20, agdcj.cid25, rrios, fssilva\}@uea.edu.br}
}

\begin{document}

\maketitle

\begin{abstract}
The growing availability of music on streaming platforms has led to information overload for users. To address this issue and enhance the user experience, increasingly sophisticated recommendation systems have been proposed. This work investigates the use of Large Language Models (LLMs) from the Gemini and LLaMA families, combined with intelligent agents, in a multi-agent personalized music recommendation system. The results are compared with a traditional content-based recommendation model, considering user satisfaction, novelty, and computational efficiency. LLMs achieved satisfaction rates of up to \textit{89{,}32\%}, indicating their promising potential in music recommendation systems.
\end{abstract}

\begin{resumo}
A crescente oferta de músicas nas plataformas de streaming tem causado sobrecarga de informação aos usuários. Para mitigar esse problema e aprimorar a experiência, sistemas de recomendação mais sofisticados têm sido propostos. Este trabalho investiga o uso de Modelos de Linguagem de Grande Escala (LLMs) das famílias Gemini e LLaMA, aliados a agentes inteligentes, em um sistema de recomendação musical personalizado multiagente. Os resultados são comparados a um modelo tradicional baseado em conteúdo, considerando satisfação do usuário, novidade e eficiência computacional. Os LLMs atingiram até \textit{89{,}32\%} de satisfação, indicando seu potencial promissor na recomendação de músicas.
\end{resumo}

\section{Introdução} \label{section:Introducao}

A música é uma das mais antigas formas de manifestação artística do ser humano. Desde os primórdios, ela desempenha um papel fundamental na expressão e na comunicação de sentimentos, emoções, opiniões, ideias etc. Devido à sua capacidade de gerar reações como, por exemplo, felicidade, tristeza ou nostalgia, torna a música um meio poderoso de conexão humana. Além disso, ela funciona como um veículo de transmissão de conhecimento, registro e preservação de histórias, permitindo que tradições e memórias sejam compartilhadas entre gerações e culturas ao longo do tempo.

Com o avanço da tecnologia e com a popularização da \textit{Internet}, a forma como as pessoas consomem música mudou radicalmente. Serviços de \textit{streaming} como, por exemplo, \textit{Spotify}\textregistered, \textit{Apple Music}\textregistered, \textit{Deezer}\textregistered e \textit{Amazon Music}\textregistered, tornaram-se as principais plataformas de distribuição musical, oferecendo aos usuários acesso rápido a catálogos com milhões de faixas. Essa grande variedade de músicas fez com que a escolha de uma música se tornasse um desafio devido à sobrecarga de informação \cite{information-overload}. Para minorar essa sobrecarga, sistemas de recomendação têm se tornado ferramentas essenciais para personalizar a experiência do usuário, facilitando assim a descoberta de músicas alinhadas aos seus gostos e preferências.

Em trabalhos recentes, diversas técnicas e mecanismos de descoberta e filtragem de músicas têm sido explorados com o intuito de melhorar a experiência do usuário e minorar a sobrecarga de informação \cite{information-overload}.  No entanto, tais abordagens enfrentam limitações importantes como, por exemplo, a dificuldade de fazer recomendações para usuários que não possuem um histórico de músicas ouvidas (\textit{cold start}) \cite{cold-start-ref}.

Nesse cenário, este trabalho propõe uma abordagem baseada em agentes inteligentes construídos sobre Modelos de Linguagem de Grande Escala (\textit{LLMs — Large Language Models}), como o \textit{Gemini} \cite{gemini} e o \textit{LLaMA} \cite{llama}, visando explorar sua capacidade de compreender linguagem natural e capturar relações contextuais e semânticas profundas. A proposta se estrutura em uma arquitetura modular composta por agentes especializados, o que favorece a flexibilidade, escalabilidade e reprodutibilidade do sistema de recomendação. Além disso, a utilização de dados reais e a avaliação com usuários finais conferem maior validade prática aos resultados obtidos. Com isso, este trabalho busca investigar não apenas as potencialidades e limitações dessa abordagem, mas também seu impacto direto na experiência do usuário em sistemas de recomendação musical.

Esse artigo está organizado como segue. A Seção \ref{section:TrabalhosRelacionados} apresenta os trabalhos relacionados. A Seção \ref{section:metodologia} detalha a metodologia utilizada no trabalho. A Seção \ref{section:Resultados} apresenta os resultados e uma discussão dos experimentos realizados. A conclusão do trabalho é apresentada na Seção \ref{section:Conclusao}.

\section{Trabalhos Relacionados}\label{section:TrabalhosRelacionados}

A recomendação musical tem sido tradicionalmente realizada por meio de abordagens em filtragem colaborativa e filtragem baseada em conteúdo. A filtragem colaborativa infere preferências a partir de padrões de comportamento semelhantes de usuários \cite{colabfiltering}. Embora eficiente, essa técnica não consegue tratar o conhecido problema do \textit{cold start}, o que dificulta a recomendação de itens para novos usuários. Para lidar com isso, alguns sistemas baseados em conteúdo utilizam metadados musicais como, por exemplo, gênero, ritmo e instrumentação, para recomendar itens de interesse do usuário \cite{schedl2014music}. Embora essa abordagem seja menos suscetível ao \textit{cold start}, ela tende a gerar recomendações mais homogêneas, o que limita a diversidade de itens recomendados.

Além dessas técnicas, métodos mais recentes incorporam aprendizado profundo para aprimorar a precisão das recomendações. Em \cite{facialemotionmodel}, é proposto um sistema de recomendação baseado na análise de emoções faciais e da emoção que a música possibilita. Esse estudo avaliou o uso de Redes Neurais Convolucionais (CNNs — \textit{Convolutional Neural Networks}) para a classificação das emoções do usuário a partir de imagens. A classificação a emoção que a música possibilita é feita utilizando Máquinas de Vetores de Suporte (SVM — \textit{Support Vector Machine}). Apesar de apresentar resultados promissores, o modelo possui limitações como, por exemplo, a dificuldade de aplicação em tempo real e a incapacidade de considerar fatores individuais, como idade, gênero e histórico de consumo do usuário.

Em \cite{vocalbasedmodel}, é apresentado um modelo de recomendação baseado em características vocais das canções. O sistema analisa o áudio das músicas para extrair informações relevantes sobre vozes e sons. As características extraídas são utilizadas como entrada de modelos convolucionais para treinamento e classificação em gêneros musicais. O resultado da classificação é combinado com o histórico de consumo do usuário para fazer as recomendações. Entretanto, o classificador proposto é ineficaz em classificar músicas em seus respectivos gêneros.

Nos últimos anos, a incorporação de LLMs aos sistemas de recomendação abriu novas possibilidades, principalmente pela capacidade dessas arquiteturas em modelar dependências contextuais e compreender nuances semânticas. Estudos como, por exemplo, \cite{BERT4Rec} e \cite{selfattentiorec} demonstraram a eficácia de arquiteturas baseadas em \textit{Transformer}, como o BERT4Rec, no sequenciamento de interações de usuários para prever preferências futuras. Embora esses métodos tenham sido inicialmente desenvolvidos para sistemas de recomendação gerais, eles inspiraram adaptações específicas no contexto musical, como o uso de \textit{embeddings} derivados de textos descritivos de faixas e artistas \cite{vagliano}.

O estudo \cite{vagliano} propõe o uso de \textit{autoencoders} adversariais para a continuação automática de \textit{playlists}. Para isso, combina múltiplas fontes de dados para enriquecer os \textit{embeddings} musicais e melhorar a qualidade das recomendações. Os resultados evidenciam que o uso de \textit{embeddings} multimodais melhora a capacidade de prever músicas adequadas para serem adicionadas a \textit{playlists}, especialmente quando as sequências são curtas ou incompletas.

Este trabalho se difere dos anteriores, especialmente, pelo uso de agentes inteligentes baseados em recomendação de músicas, comparando com o método de filtragem baseada em conteúdo amplamente utilizada em sistemas de recomendação de música.

\section{Material e Métodos} \label{section:metodologia}

Para realizar os experimentos foi feita uma coleta de dados de usuários em tempo real. Para isso, optou-se por utilizar a API pública do \textit{Spotify}\textregistered. A coleta ocorreu ao longo de 13 meses, com a participação de usuários voluntários que consentiram com o uso de seus dados para fins de pesquisa. O consentimento foi acordado considerando os princípios éticos de privacidade e confidencialidade. Após a coleta, os dados foram organizados em uma base estruturada, que contém um total de 19 usuários e 22.178 faixas únicas.

\subsection{Configuração do Ambiente de Experimentos}

Para realizar os experimentos optou-se por utilizar recursos computacionais em nuvem. A API de recomendação foi desenvolvida em Python e Django a um custo de US\$2 e foi hospedada na Amazon AWS \cite{aws} em uma instância \emph{t3.small} (2 vCPUs e 2GB RAM) e era executada em um container Docker \cite{docker}. Essa API recebe dados do MongoDB \cite{mongodb}, envia-os para agentes de IA (Inteligência Artificial) e retorna as recomendações como resultado. Os agentes utilizam os modelos LLaMA 3.3, via API da Groq, e Gemini 2.0 Flash, via API do Google. Os modelos não foram personalizados utilizando qualquer treinamento específico. As recomendações são geradas dinamicamente a cada requisição.

O \textit{front-end}, desenvolvimento em React e TypeScript e hospedado gratuitamente no Firebase, exibe as recomendações e coleta as avaliações feitas pelos usuários. O \textit{back-end}, desenvolvido em Node.js com TypeScript, atua como intermediário entre o \textit{front-end} e a API de recomendação. O \textit{back-end}, hospedado no Render a um custo de US\$ 2, gerencia as requisições recebidas do \textit{front-end} e envia para o banco de dados (MongoDB).

\subsection{Dataset}

O conjunto de dados coletados inclui dados como: nome da música, artista(s), álbum, duração da faixa, presença de conteúdo explícito, data de lançamento e data de reprodução pelo usuário. Nesse trabalho foram considerados apenas três dados: nome da música, nome do artista principal e o gênero musical. A identificação do gênero foi realizada via requisições adicionais à API do \textit{Spotify}. A base foi dividida em dois subconjuntos principais:

\begin{enumerate}
    \item \textbf{Catálogo musical:} composto por todas as faixas únicas identificadas na amostra;
    \item \textbf{Histórico musical dos usuários:} registros das músicas efetivamente ouvidas por cada participante.
\end{enumerate}

Considerando as limitações impostas pela quantidade de \textit{tokens} suportados pelos modelos de LLMs, foi realizada uma amostragem controlada. Foram selecionados os 20 gêneros musicais mais recorrentes na base original. Em seguida, utilizou-se a função \texttt{sample()} da biblioteca \texttt{Pandas} para gerar um catálogo musical com 300 faixas. Para o histórico de cada usuário, foram selecionadas as 30 músicas mais reproduzidas durante o período analisado. A Tabela \ref{tab:dataset} apresenta uma amostragem dos dados de catálogo musical.

\begin{table}[!htpb]
\centering
\caption{Amostragem do catálogo musical}
\label{tab:dataset}
% \footnotesize
\resizebox{\textwidth}{!}{%
\begin{tabular}{
    >{\raggedright\arraybackslash}p{4cm}
    >{\raggedright\arraybackslash}p{6cm}
    >{\raggedright\arraybackslash}p{6cm}
}
\hline
\textbf{Nome da Música} & \textbf{Artista (s)} & \textbf{Gêneros} \\
\hline
Decode & Sabrina Carpenter & Pop \\
\hline
Permission to Dance & BTS & K-Pop, K-Pop Boy Group, Pop \\
\hline
Bicycle Song - 2006 Remaster & Red Hot Chili Peppers & \makecell[tl]{Alternative Rock, Funk Metal,\\ Funk Rock, Permanent Wave, Rock} \\
\hline
\end{tabular}%
}
\end{table}

\subsection{Método Tradicional de Filtragem Baseada em Conteúdo}

O método tradicional, utilizado como \textit{benchmark}, consiste em um sistema de recomendação de músicas que se baseia em filtragem por conteúdo. Similarmente às implementações de \cite{fiarni2019product}, ele calcula a similaridade de cosseno entre vetores de gêneros musicais (representados por TF-IDF), uma técnica também utilizada em \cite{movie-rec-sys, rec-sys-project-supervisors}.

Inicialmente, o sistema faz requisições a uma \textit{API REST} para coletar ambos os conjuntos de dados: o histórico de músicas consumidas por um usuário; e o catálogo geral de músicas disponíveis. Em seguida, os gêneros de todas as músicas do catálogo e os gêneros preferidos do usuário são transformados em vetores numéricos por meio da técnica TF-IDF.

Com os vetores gerados, é aplicada a métrica de similaridade de cosseno (eq. \ref{eq:cos_sim}) para calcular o grau de semelhança entre os gêneros preferidos do usuário e os gêneros presentes no catálogo. O sistema seleciona as 20 músicas mais próximas aos cinco gêneros mais consumidos pelo usuário.

\begin{equation}
\text{similaridade}(x, y) = cos(\theta) = \frac{\sum_i x_i y_i}{\sqrt{\sum_i x_i^2} \sqrt{\sum_i y_i^2}}
\label{eq:cos_sim}
\end{equation}

\subsection{Método de Filtragem de Conteúdo baseado em Agentes Inteligentes}

O processo de recomendação foi conduzido por meio de agentes inteligentes baseados em LLMs, implementados com a ajuda do framework CrewAI \cite{crew}, que permite a criação de arquiteturas multiagente colaborativas. Nesse contexto, agentes inteligentes podem ser compreendidos como programas que utilizam técnicas de inteligência artificial para executar tarefas que normalmente exigiriam intervenção humana \cite{Russell2022}, sendo capazes de operar de forma autônoma e adaptativa.

Ao invés de um único \textit{prompt} genérico, a recomendação foi dividida em subtarefas executadas por agentes especializados, cada um com objetivos bem definidos como, por exemplo:

\begin{itemize}
    \item \textbf{ReadingAgt}: Leitura e compreensão do catálogo musical;
    \item \textbf{AnalistAgt}: Análise do histórico de reprodução do usuário;
    \item \textbf{ExtractAgt}: Extração de padrões de preferência com base em gêneros e artistas;
    \item \textbf{RecommendAgt}: Geração da lista final de recomendações.
\end{itemize}

\begin{figure}[ht]
    \centering
    \includegraphics[width=.55\textwidth]{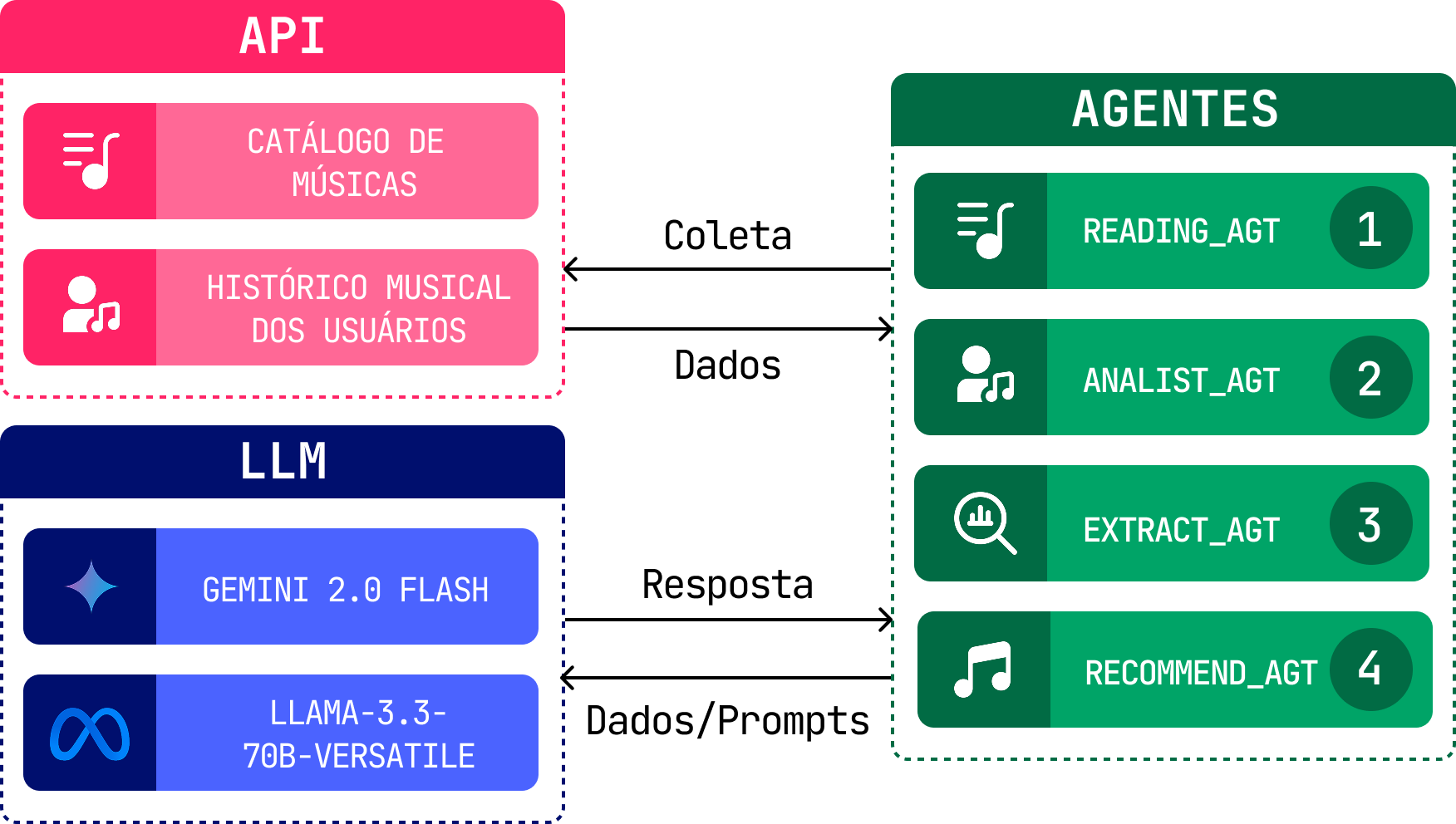}
    \caption{Fluxo do sistema de recomendação proposto}
    \label{fig-fluxo}
\end{figure}

Esses agentes foram configurados para atuar de forma independente, mas colaborativa, compartilhando informações intermediárias. Toda comunicação entre os agentes e a base de dados foi realizada via APIs REST, garantindo reprodutibilidade e modularidade no processo, conforme a Figura \ref{fig-fluxo}.

\subsubsection{Especificação dos Agentes e \textit{Prompts}}

Os \textit{prompts} empregados nos agentes seguem a abordagem \textit{zero-shot}, na qual os modelos de linguagem realizam inferências diretamente, sem a necessidade de exemplos explícitos no próprio \textit{prompt}. Esse tipo de configuração busca avaliar a capacidade dos LLMs em gerar recomendações com base na descrição textual do histórico musical do usuário e do catálogo de faixas disponíveis \cite{BERT4Rec, vagliano}.

Nas tabelas de \ref{tab:tools_table} a \ref{tab:tasks_table}, são apresentados os \textit{prompts} utilizados na construção dos agentes, bem como as ferramentas e tarefas atribuídas a cada um. A Tabela \ref{tab:tools_table} apresenta as ferramentas utilizadas pelos agentes \textbf{ReadingAgt} e \textbf{AnalistAgt}, respectivamente. Essas ferramentas foram desenvolvidas utilizando o \textit{CrewAI Tools} \cite{crew} e consistem em funções que realizam requisições à API, permitindo o retorno dos dados necessários para os agentes.

\begin{table}[!htpb]
\centering
\caption{Ferramentas utilizadas e suas respectivas descrições}
\scriptsize
\label{tab:tools_table}
\begin{tabular}{
    >{\raggedright\arraybackslash}p{4.5cm}
    >{\raggedright\arraybackslash}p{8cm}
}
\hline
\textbf{Ferramenta} & \textbf{Descrição} \\
\hline
\makecell[tl]{\texttt{GetUserHistory}\\\texttt{DataTool}} & \textit{Fetches the music data from a given URL and returns it as a list of dictionaries.} \\
\hline
\makecell[tl]{\texttt{GetMusicCatalogue}\\\texttt{Tool}} & \textit{Fetches the user listening history from a given URL and returns the first 30 items as a list.} \\
\hline
\end{tabular}
\end{table}

A Tabela \ref{tab:agents_table} apresenta os agentes previamente introduzidos, acompanhados de suas respectivas descrições de objetivo e contexto. A descrição de objetivo serve como diretriz principal, orientando as decisões e ações de cada agente ao longo do processo de recomendação. O contexto fornece informações adicionais que moldam a personalidade, o comportamento e a forma de atuação dos agentes \cite{crew}.

\begin{table}[!htpb]
\centering
\caption{Agentes utilizados, objetivos e ferramentas associadas}
\footnotesize
\label{tab:agents_table}
\resizebox{\textwidth}{!}{%
\begin{tabular}{
    >{\raggedright\arraybackslash}p{3.5cm}
    >{\raggedright\arraybackslash}p{5cm}
    >{\raggedright\arraybackslash}p{5cm}
    >{\raggedright\arraybackslash}p{3cm}
}
\hline
\textbf{Agente} & \textbf{Objetivo (Goal)} & \textbf{Contexto (Backstory)} & \textbf{Ferramenta (Tool)} \\
\hline
\texttt{\textbf{ReadingAgt}} & \textit{Read all the songs from a catalogue.} & \textit{Specializes in handling and returning a song catalogue.} & \makecell[tl]{\texttt{GetMusic}\\\texttt{CatalogueTool}} \\
\hline
\texttt{\textbf{AnalistAgt}} & \textit{Read all a music history from an user.} & \textit{Specializes in handling and returning a music history.} & \makecell[tl]{\texttt{GetUserHistory}\\\texttt{DataTool}} \\
\hline
\texttt{\textbf{ExtractAgt}} & \textit{Inferring the user's favorite music genre.} & \textit{Specializes in analysing the user music history to infer their 5 favorite music genres.} & - \\
\hline
\texttt{\textbf{RecommendAgt}} & \textit{Recommend songs to a user using their listening histories.} & \textit{You are a personalized music recommender. You analyze song genres and recommend tracks using content-based filtering techniques.} & - \\
\hline
\end{tabular}%
}
\end{table}

A Tabela \ref{tab:tasks_table} apresenta a estruturação dos \textit{prompts} utilizados nas tarefas, organizados de forma a guiar a execução passo a passo do processo de recomendação. Essa organização visa garantir clareza, coesão e eficiência na atuação de cada agente dentro do sistema. As tarefas, por sua vez, correspondem a atribuições específicas designadas aos agentes, contendo todas as informações necessárias para sua execução, como uma descrição detalhada, o agente responsável, as ferramentas requeridas, entre outros elementos essenciais que viabilizam a correta execução das ações \cite{crew}.

\begin{table}[!htpb]
\centering
\caption{Tarefas utilizadas no sistema}
\label{tab:tasks_table}
\small
\resizebox{\textwidth}{!}{%
\begin{tabular}{
    >{\raggedright\arraybackslash}p{8cm}
    >{\raggedright\arraybackslash}p{5cm}
    >{\raggedright\arraybackslash}p{4cm}
}
\hline
\textbf{Descrição} & \textbf{Agente} & \textbf{Saída Esperada} \\
\hline
\textit{Read and return all the song catalogue at this URL:} \texttt{/getAllDataEniac?limit=300}. & \texttt{Song Catalogue Reader} & \textit{Song Catalogue} \\
\hline
\textit{Read and return the user music history at this URL:} \texttt{/getUserData/\{user\_id\}} & \texttt{User Music History Reader} & \textit{User listening history}  \\
\hline
\textit{Infer the user's favorite music genres. Use the user's listening history to identify their 5 most preferred music genres.} & \texttt{User Music Genres} & \textit{User Music Genres} \\
\hline
\textit{Generate a list of 20 recommended songs from the catalogue. Use the user's listening history and inferred genres to build a personalized profile. Select songs that match the user's musical preferences. Return a JSON list with \texttt{genre}, \texttt{song\_name} and \texttt{artist\_name}, along with \texttt{liked} and \texttt{known} flags.} & \texttt{Content-Based Music Recommender} & \textit{Recommended Songs for the user}  \\
\hline
\end{tabular}%
}
\end{table}

\subsubsection{Seleção dos Modelos}

Para a implementação do sistema de recomendação com agentes baseados em LLMs, a seleção dos modelos subjacentes foi guiada por critérios de capacidade técnica e viabilidade operacional, especialmente no contexto de acesso e custo. Foram avaliados o Gemini 2.0 Flash \cite{google2025gemini_2_0_flash} e o LLaMA-3.3-70B-VERSATILE  \cite{groq} como os LLMs principais para as inferências dos agentes.

O Gemini 2.0 Flash foi escolhido principalmente devido à sua capacidade de processar um elevado número de \textit{tokens} em sua versão gratuita, oferecendo uma janela de contexto de até 1 milhão de \textit{tokens}. Essa característica se mostrou decisiva para o projeto, pois permitiu o processamento de catálogos musicais extensos e históricos de reprodução detalhados dos usuários sem incorrer em custos adicionais significativos durante a fase de desenvolvimento e experimentação.

O modelo LLaMA-3.3-70B-VERSATILE foi escolhido devido à sua facilidade de acesso e custo-benefício através do serviço sob demanda da plataforma \emph{Groq}, na qual permite o uso de LLMs aceleradas por meio de acesso de API \cite{groq}. Apesar de não apresentar uma janela de \textit{tokens} tão elevada quanto a versão gratuita do Gemini Flash, o custo acessível do serviço viabilizou a utilização de um modelo de grande porte (70 bilhões de parâmetros) para as operações de inferência mais complexas como, por exemplo, a análise de histórico e a inferência de padrões de preferência. A escolha de um modelo com 70B de parâmetros subentende a busca por uma maior capacidade de compreensão e geração de linguagem natural, permitindo aos agentes capturar nuances contextuais e semânticas mais profundas necessárias para uma recomendação musical sofisticada, mesmo que a velocidade e o custo tenham sido fatores mais diretos na decisão de acesso via \emph{Groq}.

Ambos os modelos foram acessados por meio de chaves de API, o Gemini foi acessado via API proprietária, enquanto o LLaMA foi acessado via serviço sob demanda oferecido pela plataforma \emph{Groq}.

\subsection{Método de Avaliação}

Foi desenvolvida uma interface específica para que os usuários interagissem com as \textit{playlists} geradas pelos diferentes métodos de recomendação. Cada participante teve acesso às três \textit{playlists}, compostas com 10 faixas recomendadas, correspondentes aos métodos avaliados (tradicional, LLaMA e Gemini), de forma cega, e realizou a avaliação de cada uma individualmente. A análise foi conduzida com base nos seguintes critérios:

\begin{enumerate}
    \item Se gostou da música recomendada (resposta binária: Gostei / Não gostei);
    \item Se já conhecia a faixa apresentada (resposta binária: Sim / Não);
    \item Avaliação geral da \textit{playlist}, em uma escala ordinal de 0 a 10, sendo 0 equivalente a "péssimo" e 10 a "ótimo".
\end{enumerate}

Com as respostas coletadas, são calculadas as seguintes métricas de avaliação, as quais serão usadas para comparação de desempenho dos diferentes modelos na tarefa de recomendação de músicas.

\paragraph{Taxa de Apreciação (\textit{Like Rate}).}
Define-se como a proporção de músicas curtidas pelo usuário:

\begin{equation} \label{eq:LikeRate}
    \text{LR} = \frac{1}{N} \sum_{i=1}^{N} \mathbb{I}(\text{like}_i = 1)
\end{equation}

onde $N$ é o número total de músicas avaliadas pelo usuário na \textit{playlist}, e $\mathbb{I}$ é a função indicadora.

\paragraph{Taxa de Descoberta (\textit{Novelty Rate}).}
Corresponde à fração de músicas não previamente conhecidas pelo usuário:

\begin{equation} \label{eq:NR}
    \text{NR} = \frac{1}{N} \sum_{i=1}^{N} \mathbb{I}(\text{known}_i = 0)
\end{equation}

\paragraph{Taxa de Descobertas Bem-Sucedidas (\textit{Successful Novelty Rate}).}
Proporção de músicas \emph{novas} que foram curtidas, refletindo a eficácia da recomendação de faixas inéditas:

\begin{equation} \label{eq:SNR}
    \text{SNR} =
    \frac{\sum_{i=1}^{N} \mathbb{I}(\text{known}_i = 0 \wedge \text{like}_i = 1)}
    {\sum_{i=1}^{N} \mathbb{I}(\text{known}_i = 0)}
\end{equation}

Essa métrica é condicional à existência de pelo menos uma música nova na \textit{playlist}.

\paragraph{Nota da Playlist (\textit{Playlist Rating}).}
Refere-se à nota atribuída pelo usuário à \textit{playlist} como um todo, após ouvir as dez faixas. Essa é uma medida subjetiva global da qualidade percebida da recomendação.

As métricas descritas permitem avaliar não apenas a eficácia da recomendação em termos de apreciação geral, mas também sua capacidade de promover descobertas relevantes aos usuários.

\section{Resultados e Discussão} \label{section:Resultados}

\begin{figure}[ht]
    \centering
    \includegraphics[width=.55\textwidth]{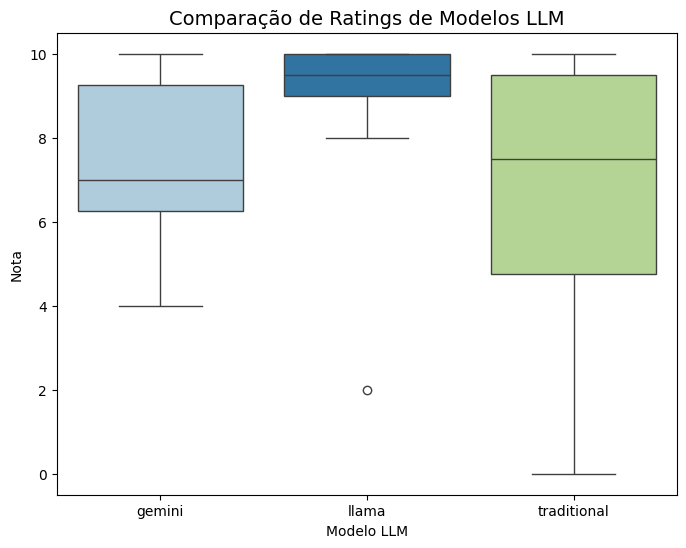}
    \caption{Boxplot da Comparação de Ratings dos Modelos LLM}
    \label{fig-recsys}
\end{figure}

Os experimentos seguiram o protocolo descrito na Seção \ref{section:metodologia}. A Tabela \ref{tab:models_metrics} e a Figura~\ref{fig-recsys} apresentam os resultados médios das avaliações feitas por dez usuários, de forma cega, expressos em termos de média e desvio padrão. Considerando exclusivamente a métrica de \textit{Rating}, que reflete a avaliação subjetiva global da qualidade da playlist, observa-se uma diferença clara entre os modelos.

\begin{table}[!htpb]
\centering
\caption{Métricas de avaliação dos modelos em termos de média e desvio padrão.}
\label{tab:models_metrics}
\resizebox{\textwidth}{!}{%
\begin{tabular}{lccccc}
\hline
\textbf{Modelo} & \textbf{LR (\%)} & \textbf{NR (\%)} & \textbf{SNR (\%)} & \textbf{Rating (1--10)} & \textbf{Tempo de Inferência (s)} \\
\hline
Tradicional & $61{,}00 \pm 22{,}00$ & $58{,}50 \pm 22{,}48$ & $21{,}00 \pm 18{,}55$ & $6{,}70 \pm 3{,}37$ & $1{,}37 \pm 0{,}28$ \\
LLaMA       & $89{,}32 \pm 7{,}34$ & $11{,}85 \pm 8{,}76$ & $3{,}17 \pm 3{,}61$ & $8{,}70 \pm 2{,}45$ & $84{,}07 \pm 13{,}84$ \\
Gemini      & $65{,}00 \pm 21{,}10$  & $52{,}00 \pm 23{,}79$ & $18{,}50 \pm 18{,31}$ & $7{,}25 \pm 2{,}18$ & $70{,}76 \pm 32{,}80$ \\
\hline
\end{tabular}%
}
\end{table}

A análise das métricas de desempenho revelou padrões distintos entre os modelos LLaMA, Gemini e o Tradicional, especialmente no que tange à satisfação do usuário e à capacidade de descoberta. O modelo \textbf{LLaMA} obteve a maior média de \textit{Rating}, com $8{,}70 \pm 2{,}45$, indicando uma forte aceitação por parte dos participantes. Esse desempenho sugere que suas recomendações foram percebidas como mais coerentes, agradáveis ou relevantes em comparação às demais. Em segundo lugar, o modelo \textbf{Gemini} alcançou média de $7{,}25 \pm 2{,}18$, o que também representa uma avaliação positiva, embora com leve inferioridade em relação ao LLaMA.

Por outro lado, o modelo \textbf{Tradicional} apresentou a menor média de nota, com $6{,}70 \pm 3{,}37$. Além da menor média, o desvio padrão mais elevado sugere uma maior variabilidade na percepção dos usuários sobre suas \textit{playlists}, indicando que, embora algumas tenham sido bem recebidas, outras foram avaliadas de forma mais crítica.

O modelo \textbf{LLaMA} também obteve a maior taxa de apreciação ($89{,}32\% \pm 7{,}34$), sugerindo que, mesmo com baixa novidade, suas recomendações estavam altamente alinhadas às preferências dos usuários. Isso pode indicar uma forte capacidade de personalização, porém possivelmente com tendência a recomendações conservadoras, já conhecidas ou semelhantes a preferências passadas. O \textbf{Gemini}, por sua vez, obteve uma taxa de apreciação intermediária ($65{,}00\%$), alinhando-se mais ao desempenho do modelo \textbf{Tradicional} ($61{,}00\%$), mas ainda com melhor aceitação geral da \textit{playlist}.

Peculiarmente, os dados mostram uma inversão entre novidade e apreciação. O modelo \textbf{LLaMA} apresentou a menor taxa de descoberta ($11{,}85\%$), o que pode indicar que a maioria das faixas já era conhecida pelos usuários. Em contrapartida, o modelo \textbf{Tradicional} atingiu a maior taxa de descoberta ($58{,}50\%$), seguido pelo Gemini ($52{,}00\%$). Isso sugere que, apesar de mais inovadores, esses modelos tiveram menor alinhamento com as preferências dos usuários, o que pode estar relacionado às suas avaliações mais baixas.

A métrica de descoberta bem-sucedida (SNR) revelou diferenças significativas entre os modelos. O \textbf{Tradicional} se destacou com a maior SNR ($21{,}00\%$), demonstrando sua eficácia em apresentar músicas inovadoras e apreciadas, quando comparado aos demais modelos. Esse desempenho contrasta fortemente com a do \textbf{LLaMA}, que registrou a menor SNR ($3{,}17\%$), confirmando a hipótese de que esse modelo tende a focar em faixas já conhecidas, limitando sua capacidade de exploração. O \textbf{Gemini}, por sua vez, atingiu um equilíbrio com uma SNR de $18{,}50\%$. Isso indica que ele é mais apto a introduzir novas músicas agradáveis do que o \textbf{LLaMA}, mas ainda não alcança o desempenho do \textbf{Tradicional} nesse quesito.

Além das métricas de satisfação e descoberta, a eficiência computacional dos modelos, medida pelo tempo de inferência, revelou diferenças significativas que impactam diretamente sua aplicabilidade. O modelo \textbf{Tradicional} demonstrou ser o mais rápido e eficiente, com um tempo de inferência de apenas $1{,}37 \pm 0{,}28$ seg. Essa velocidade o torna mais adequado para cenários que exigem respostas em tempo real ou processamento eficiente de grandes volumes de dados. Em contrapartida, os modelos baseados em LLMs apresentaram tempos de inferência substancialmente mais longos. O \textbf{Gemini} registrou um tempo de $70{,}76 \pm 32{,}80$ seg., enquanto o \textbf{LLaMA} foi o mais lento, com $84{,}07 \pm 13{,}84$ seg. Essa diferença de magnitude em relação ao \textbf{Tradicional} é um fator crítico. Embora o \textbf{LLaMA} tenha se destacado em termos de satisfação do usuário e apreciação, o tempo de resposta pode ser um gargalo em sistemas que necessitam de baixa latência.

\section{Conclusão e Trabalhos Futuros} \label{section:Conclusao}

Este trabalho investigou e comparou o desempenho de modelos de recomendação de \textit{playlists}, Tradicional, LLaMA e Gemini, sob as perspectivas de satisfação do usuário, capacidade de descoberta de novas músicas e eficiência computacional. Os resultados obtidos revelam um cenário complexo, onde cada abordagem apresenta vantagens e limitações distintas.

Embora os LLMs selecionados (LLaMA e Gemini) tenham a capacidade de gerar recomendações altamente personalizadas e apreciadas pelos usuários (altas notas de \textit{Rating} do LLaMA e sua elevada taxa de apreciação), eles tendem a ser conservadores em termos de taxa de novidade. Por outro lado, o modelo Tradicional, embora com menores índices de satisfação geral, provou ser mais eficaz na introdução de descobertas bem-sucedidas (SNR). Isso sugere que, para aplicações onde a curadoria e a introdução de material novo são prioridades, as abordagens tradicionais ainda podem ser mais vantajosas.

Um ponto crítico revelado na avaliação dos modelos foi a eficiência computacional. Apesar de apresentar melhores resultados quanto à satisfação do usuário, os LLMs apresentam uma baixa eficiência computacional, contrastando com o modelo Tradicional, que opera em baixa latência. Em ambientes que exigem respostas em tempo real, a superioridade dos LLMs em termos de personalização pode ser ofuscada por suas limitações de desempenho, tornando-os menos viáveis nessa tarefa. Para sistemas que priorizam a satisfação imediata do usuário e a personalização de preferências já estabelecidas e que toleram maior tempo de resposta, modelos como o LLaMA mostram-se promissores. Para cenários que valorizam a descoberta genuína de novos conteúdos e a eficiência computacional, as abordagens tradicionais ainda mantêm sua relevância. O Gemini, por sua vez, posiciona-se como um intermediário, pois equilibra personalização e alguma capacidade de descoberta, embora ainda com um custo computacional elevado.

Como trabalhos futuros, planejamos aprimorar a personalização das recomendações musicais por meio do uso de um maior volume de dados de catálogo e histórico musical, além da ampliação dos atributos utilizados como entrada dos modelos. Pretendemos ir além de informações básicas como nome da música, artista e gênero, incorporando também letras, sentimentos extraídos, descrições contextuais e outros elementos que possam explorar melhor as capacidades dos LLMs. Para isso, investigaremos abordagens híbridas de filtragem, combinando técnicas baseadas em conteúdo e filtragem colaborativa. Para avaliar o uso de catálogos e históricos de consumo maiores, pretendemos utilizar bancos de dados vetoriais como suporte à arquitetura do sistema e avaliar seu impacto no desempenho dos LLMs nesse cenário. Pretende-se também utilizar outros modelos como, por exemplo, Gemma, Mistral e Qwen. Por fim, será conduzida uma análise do efeito de diferentes valores do hiperparâmetro de temperatura nas respostas dos modelos, para avaliar se esse fator influencia a qualidade, diversidade e relevância das recomendações geradas.

%% Agradecimentos

\bibliographystyle{sbc}
\bibliography{main}

\begin{thebibliography}{}

\bibitem[AWS 2025]{aws}
AWS (2025).
\newblock Amazon web services.
\newblock \url{https://www.aws.amazon.com/}.
\newblock Acesso em: 13 mai. 2025.

\bibitem[CrewAI 2025]{crew}
CrewAI (2025).
\newblock The leading multi-agent platform.
\newblock \url{https://www.crewai.com/}.
\newblock Acesso em: 30 jun. 2025.

\bibitem[Docker 2025]{docker}
Docker (2025).
\newblock Docker: Accelerated container application development.
\newblock \url{https://www.docker.com/}.
\newblock Acesso em: 30 jun. 2025.

\bibitem[Falah and Suryawan 2022]{rec-sys-project-supervisors}
Falah, Z.~F. and Suryawan, F. (2022).
\newblock Recommendation system to propose final project supervisors using
  cosine similarity matrix.
\newblock {\em Khazanah Informatika Jurnal Ilmu Komputer dan Informatika},
  8(2).

\bibitem[Fiarni and Maharani 2019]{fiarni2019product}
Fiarni, C. and Maharani, H. (2019).
\newblock Product recommendation system design using cosine similarity and
  content-based filtering methods.
\newblock {\em IJITEE (International Journal of Information Technology and
  Electrical Engineering)}, 3(2):42--48.

\bibitem[Fouad et~al. 2025]{information-overload}
Fouad, O., Fouad, R., Hussen, N., and Abuhadrous, I. (2025).
\newblock A comprehensive review of music recommendation systems.
\newblock {\em Adv. Sciences and Technology Journal}, 2(1):1--18.

\bibitem[{Gemini Team and Google DeepMind} 2023]{gemini}
{Gemini Team and Google DeepMind} (2023).
\newblock {Gemini: A Family of Highly Capable Multimodal Models}.
\newblock {\em arXiv preprint arXiv:2312.11805}.
\newblock Acesso em: 5 jun. 2025.

\bibitem[{Google} 2025]{google2025gemini_2_0_flash}
{Google} (2025).
\newblock {Gemini 2.0: Flash, Flash-Lite and Pro - Google Developers Blog}.
\newblock \url{https://developers.googleblog.com/en/gemini-2-family-expands/}.
\newblock Acesso em: 28 mai.2025.

\bibitem[Grattafiori et~al. 2024]{llama}
Grattafiori, A., Dubey, A., Jauhri, A., Pandey, A., Kadian, A., Al-Dahle, A.,
  and et~al. (2024).
\newblock The llama 3 herd of models.

\bibitem[Groq 2025]{groq}
Groq (2025).
\newblock Groq is fast ai inference.
\newblock \url{https://www.groq.com/}.
\newblock Acesso em: 30 jun. 2025.

\bibitem[Kang and McAuley 2018]{selfattentiorec}
Kang, W.-C. and McAuley, J. (2018).
\newblock Self-attentive sequential recommendation.
\newblock In {\em 2018 IEEE International Conference on Data Mining (ICDM)},
  pages 197--206.

\bibitem[Maheshwari 2023]{cold-start-ref}
Maheshwari, C. (2023).
\newblock Music recommendation on spotify using deep learning.

\bibitem[MongoDB 2025]{mongodb}
MongoDB (2025).
\newblock Mongodb: The world's leading modern database | mongodb.
\newblock \url{https://www.mongodb.com/}.
\newblock Acesso em: 30 jun. 2025.

\bibitem[Nguyen et~al. 2024]{facialemotionmodel}
Nguyen, H., Tran, N., Ly, D., Tran, A., Nguyen, A., Vo, H., et~al. (2024).
\newblock A model for song recommendation based on facial emotion analysis and
  musical emotion.
\newblock {\em International Journal of Intelligent Engineering \& Systems},
  17(4).

\bibitem[Russell and Norvig 2022]{Russell2022}
Russell, J. and Norvig, P. (2022).
\newblock {\em Artificial Intelligence - A Modern Approach}.
\newblock GEN LTC, 4th edition.

\bibitem[Sarwar et~al. 2001]{colabfiltering}
Sarwar, B., Karypis, G., Konstan, J., and Riedl, J. (2001).
\newblock Item-based collaborative filtering recommendation algorithms.
\newblock In {\em Proceedings of the 10th International Conference on World
  Wide Web}, WWW '01, page 285–295, New York, NY, USA. Association for
  Computing Machinery.

\bibitem[Schedl et~al. 2014]{schedl2014music}
Schedl, M., G{\'o}mez, E., Urbano, J., et~al. (2014).
\newblock Music information retrieval: Recent developments and applications.
\newblock {\em Foundations and Trends{\textregistered} in Information
  Retrieval}, 8(2-3):127--261.

\bibitem[Singh et~al. 2020]{movie-rec-sys}
Singh, R.~H., {Inderprastha Engineering College, AKTU}, Maurya, S., Tripathi,
  T., Narula, T., Srivastav, G., {Inderprastha Engineering College, AKTU},
  {Inderprastha Engineering College, AKTU}, {Inderprastha Engineering College,
  AKTU}, and {Inderprastha Engineering College, AKTU} (2020).
\newblock Movie recommendation system using cosine similarity and {KNN}.
\newblock {\em Int. J. Eng. Adv. Technol.}, 9(5):556--559.

\bibitem[Sun et~al. 2019]{BERT4Rec}
Sun, F., Liu, J., Wu, J., Pei, C., Lin, X., Ou, W., and Jiang, P. (2019).
\newblock Bert4rec: Sequential recommendation with bidirectional encoder
  representations from transformer.
\newblock In {\em Proceedings of the 28th ACM International Conference on
  Information and Knowledge Management}, CIKM '19, page 1441–1450, New York,
  NY, USA. Association for Computing Machinery.

\bibitem[Vagliano et~al. 2018]{vagliano}
Vagliano, I., Galke, L., Mai, F., and Scherp, A. (2018).
\newblock Using adversarial autoencoders for multi-modal automatic playlist
  continuation.
\newblock In {\em Proceedings of the ACM Recommender Systems Challenge 2018},
  RecSys Challenge '18, New York, NY, USA. Association for Computing Machinery.

\bibitem[Yang 2022]{vocalbasedmodel}
Yang, J. (2022).
\newblock Personalized song recommendation system based on vocal
  characteristics.
\newblock {\em Mathematical Problems in Engineering}, 2022(1):3605728.

\end{thebibliography}

\end{document}